\documentclass{mn2e}
\usepackage{epsfig}

\title[A Census of the Carina Nebula]{A Census of the Carina Nebula --
II. Energy Budget and Global Properties of the Nebulosity}

\author[N.\ Smith \& K.J.\ Brooks]{Nathan
Smith$^1$\thanks{nathans@astro.berkeley.edu} and Kate J.\
Brooks$^2$\thanks{Kate.Brooks@csiro.au} \\ $^1$Astronomy Department,
University of California, 601 Campbell Hall, Berkeley, CA 94720, USA
\\ $^2$Australia Telescope National Facility, CSIRO, PO Box 76,
Epping, NSW 1710, Australia}

\date{Accepted 0000, Received 0000, in original form 0000}
\pagerange{\pageref{firstpage}--\pageref{lastpage}} \pubyear{2002}

\def\arcdeg{\degr}

\begin{document}
\label{firstpage}
\maketitle
\begin{abstract}

The first paper in this series took a direct census of energy input
from the known OB stars in the Carina Nebula, and in this paper we
study the global properties of the surrounding nebulosity.  This
detailed comparison may prove useful for interpreting observations of
extragalactic giant H~{\sc ii} regions and ultraluminous infrared
galaxies.  We find that the total IR luminosity of Carina is about
1.2$\times$10$^7$ L$_{\odot}$, accounting for only about 50--60\% of
the known stellar luminosity from Paper I.  Similarly, the ionizing
photon luminosity derived from the integrated radio continuum is about
7$\times$10$^{50}$ s$^{-1}$, accounting for $\sim$75\% of the expected
Lyman continuum from known OB stars.  The total kinetic energy of the
nebula is about 8$\times$10$^{51}$ ergs, or $\sim$30\% of the
mechanical energy from stellar winds over the lifetime of the nebula,
so there is no need to invoke a supernova (SN) explosion based on
energetics.  Warm dust grains residing in the H~{\sc ii} region
interior dominate emission at 10-30 $\mu$m, but cooler grains at
30--40~K dominate the IR luminosity and indicate a likely gas mass of
$\sim$10$^6$ M$_{\odot}$.  We find an excellent correlation between
the radio continuum and 20-25 $\mu$m emission, consistent with the
idea that the $\sim$80 K grain population is heated by trapped
Ly$\alpha$ photons.  Similarly, we find a near perfect correlation
between the far-IR optical depth map of cool grains and 8.6~$\mu$m
hydrocarbon emission, indicating that most of the nebular mass resides
as atomic gas in photodissociation regions and not in dense molecular
clouds.  Synchronized star formation around the periphery of Carina
provides a strong case that star formation here was indeed triggered
by stellar winds and UV radiation. This second generation appears to
involve a cascade toward preferentially intermediate- and low-mass
stars, but this may soon change when $\eta$~Car and its siblings
explode.  If the current reservoir of atomic and molecular gas can be
tapped at that time, massive star formation may be rejuvinated around
the periphery of Carina much as if it were a young version of Gould's
Belt.  Also, when these multiple SNe occur, the triggered second
generation will be pelted repeatedly with SN ejecta bearing
short-lived radioactive nuclides.  Carina may therefore represent the
most observable analog to the cradle of our own Solar System.

\end{abstract}


\section{INTRODUCTION}

The Carina Nebula (NGC3372) is the nearest massive star-forming region
that satisfies three criteria which distinguish it from other Galactic
giant H~{\sc ii} regions: 1) it harbors the most extreme grouping of
massive stars within a few kpc of the Sun, including $\sim$70 O stars
initially (some, like $\eta$ Car, have now moved off the main
sequence) that are in the process of creating a giant superbubble, 2)
it is young enough that active star formation is still ongoing within
a few parsecs of these massive stars, and 3) unlike any comparable
massive cluster in our Galaxy, our sightline to it suffers little
interstellar extinction, allowing it to be studied across the
electromagnetic spectrum and not just at infrared (IR) and radio
wavelengths.  This last point is critical, because it offers the
potential to construct a relatively complete picture of the massive
star-forming environment and the stellar content.  The first two
indicate that Carina is our nearest suitable analog of more extreme
regions like 30 Doradus.  Finally, unlike many Galactic regions, the
distance to Carina (2.3 kpc) is known to within a few percent from the
expansion parallax of $\eta$ Carinae's circumstellar nebula (see Smith
2006b, 2002a; Allen \& Hillier 1993), so that the properties one
derives are reliable.

In this paper we aim to assess the total energy budget of the region
and the global properties of the large-scale nebulosity.  In a
previous paper (Smith 2006a; hereafter Paper I) we compiled the total
energy input from the known stellar population, including both
radiation and mechanical energy from stellar winds.  Thus, with the
energy input known, we can evaluate the efficiency with which stellar
radiation is reprocessed by the gas and dust.  Carina is an ideal
laboratory for comprehensive multiwavelength studies of an environment
much like that where our own solar system may have formed, where young
protoplanetary disks will be bombarded by supernova ejecta.  It also
represents the early stages of the birth of an OB association, and it
is an environment where this young OB association is triggering the
birth of a second generation of stars as they destroy their own natal
giant molecular cloud.

Our goals here are twofold.  First, we wish to evaluate the global
energy budget of the entire Carina Nebula by examining its overall,
average observed properties, much as if it were an unresolved
extragalactic giant H~{\sc ii} region.  We can then compare this
nebular energy budget with the known stellar energy input (Paper I).
This provides a check on the validity of deriving stellar properties
using similar observations of more distant H~{\sc ii} regions where
the stars cannot be counted individually (e.g., Kennicutt 1984, 1998),
as well as the most extreme examples of star formation in
ultraluminous infrared galaxies (ULIRGs).  Second, we examine
large-scale correlations between various tracers of ionized gas,
polycyclic aromatic hydrocarbons (PAHs), and warm and cold dust
properties, providing more complete insight to the global structure of
giant H~{\sc ii} regions.

\section{MULTIWAVELENGTH OBSERVATIONS}

\subsection{Visual-Wavelength Images}

We incorporate wide-field optical images of the Carina region taken
through narrowband filters transmitting [O~{\sc iii}] $\lambda$5007,
H$\alpha$, and [S~{\sc ii}] $\lambda\lambda$6717,6731, as well as
images through the broadband $B$ and $R$ filters.  The images were
already presented and discussed briefly by Smith et al.\ (2000), and
were obtained in 1999 April as part of the Mount Stromlo and Siding
Spring Observatory (MSSSO) H$\alpha$ survey (Buxton et al.\ 1998).
See Smith et al.\ (2000) for more details.  Figure 1$a$ shows a
three-color version of the [O~{\sc iii}] + H$\alpha$ + [S~{\sc ii}]
image.  The H$\alpha$ filter is contaminated by emission from [N~{\sc
ii}] $\lambda$6583, but this effect is less than 5\% over the
brightest parts of the nebula (Smith et al.\ 2004b).  While the
[N~{\sc ii}]/H$\alpha$ ratio rises in ionization fronts at the edges
of the nebula, these faint outer regions contribute little to the
total H$\alpha$ flux.  The narrowband images used here were flux
calibrated in ergs s$^{-1}$ cm$^{-2}$ arcsec$^{-2}$ (integrated over
each filter bandwidth) by matching counts in diffuse regions near the
Keyhole nebula to the observed surface brightness in the same regions
in flux-calibrated images of the Keyhole obtained with {\it HST}/WFPC2
(see Smith et al.\ 2004b).  The calibrated images have been corrected
for an average interstellar extinction of $E(B-V)$=0.4 and $R$=4.8
indicated by nebular gas in that part of the inner Carina Nebula
(Smith 2002b; Smith et al.\ 2004b), but obviously this does not
correct for patches of local extinction by dark clouds.

\subsection{Mid-IR:  MSX}

The Carina Nebula was observed by the {\it Midcourse Space Experiment}
({\it MSX}) satellite, which is a Ballistic Missile Defense
Organization satellite launched in 1996 April, equipped with a 33cm
diameter telescope and IR imager called SPIRIT III.  Details of the
instrumentation are given by Egan et al.\ (1998), while a description
of the astronomical experiments is given by Price (1995).  A brief
initial discussion of the {\it MSX} observations of Carina was already
given by Smith et al.\ (2000), but a more thorough analysis will be
undertaken here.  We consider observations of Carina obtained in four
filters: Band A ($\Delta\lambda$=6.8--10.8 $\micron$), Band C
($\Delta\lambda$=11--15.3 $\micron$), Band D ($\Delta\lambda$=13--16.5
$\micron$), and Band E ($\Delta\lambda$=17.5--27.5 $\micron$).  In
general, for extended regions, Band A is dominated by PAH emission
features, while Band E is dominated by thermal continuum emission from
hot dust grains.  Band E also includes a broad 21.3~$\micron$ emission
feature, presumably caused by silicate emission, which is strong in
the inner Carina Nebula (Chan \& Onaka 2000) and the supernova remnant
Cas A (Arendt et al.\ 1999).  The effective spatial resolution in
these images is about 18\arcsec\ or better.

Several individual image tiles for each filter were obtained from the
archive at the Infrared Processing and Analysis Center (IPAC) and
mosaiced together to produce a wide-field image of the entire region.
The resulting mosaic images in each filter were then spatially aligned
with the optical images, resulting in a pixel scale of 12$\farcs$35
and a total field of view of 2.8$\times$4 square degrees, as shown in
Figure 1$b$.  The original image data were calibrated in irradiance
units of W m$^{-2}$ ster$^{-1}$.  In order to measure the total flux
from the region, we converted these to Jy arcsec$^{-2}$ using the
bandwidth of each filter.

\subsection{Far-IR:  IRAS}

As with the {\it MSX} data, we used the IPAC archive to retrieve
several tiled images obtained by the {\it Infrared Astronomical
Satellite} ({\it IRAS}).  We mosaiced together the individual tiles
into a large image in each of the four {\it IRAS} bands (12, 25, 60,
and 100 $\micron$), and then spatially aligned these to the optical
and {\it MSX} images.  The result is shown in Figure 1$c$.  We did not
use higher-resolution {\it IRAS} images at 60 and 100 $\micron$ from
the {\it IRAS} Glaxy Atlas because we are most interested in global
properties, and we need accurate absolute fluxes to investigate dust
color temperatures and optical depths from image ratios at various
wavelengths.

\begin{figure*}\begin{center}
\epsfig{file=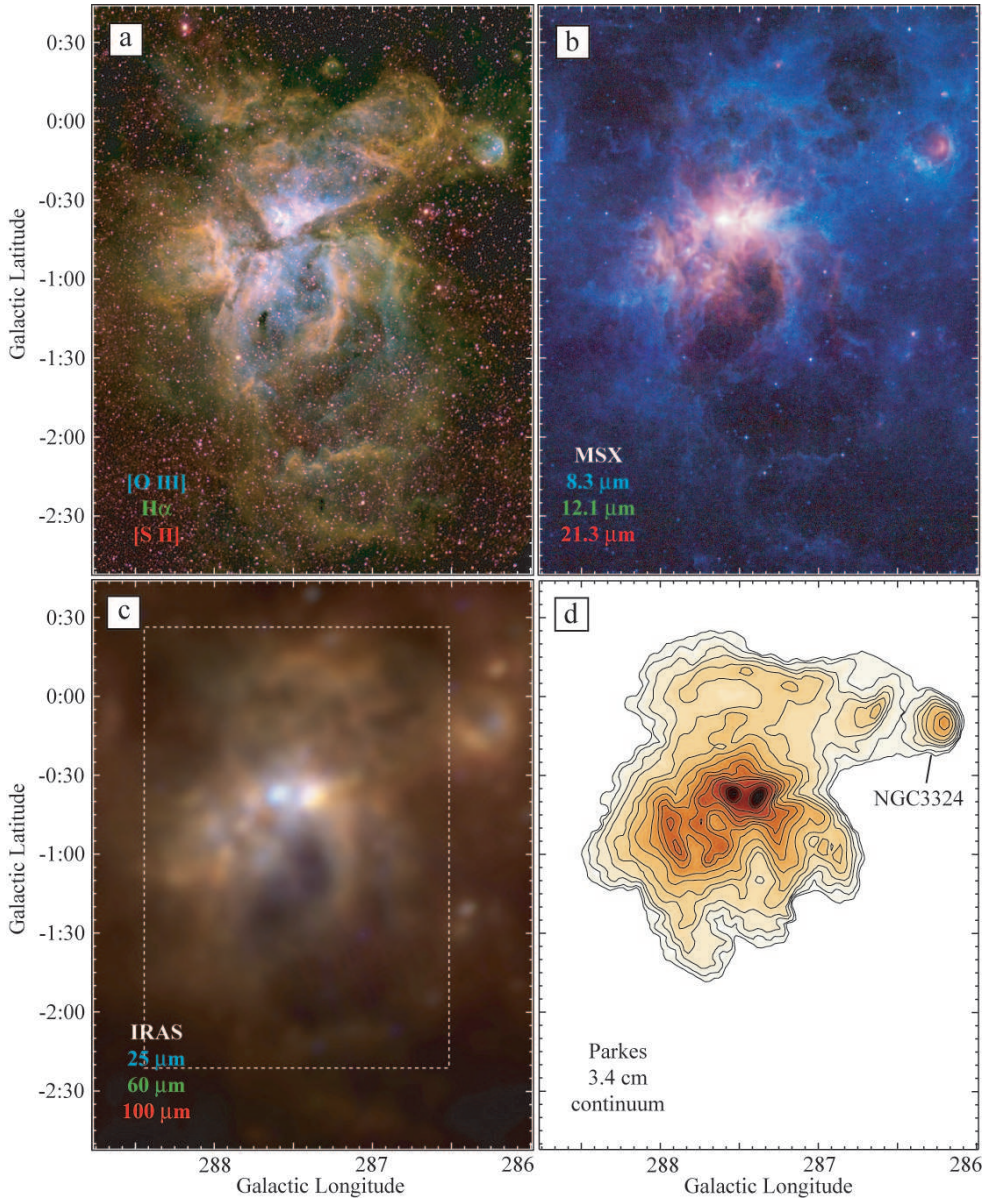,width=5.1in}
\end{center}
\caption{Wide field images of the entire Carina Nebula in the optical,
IR, and radio.  (a) Composite color image with blue = [O~{\sc iii}]
$\lambda$5007, green = H$\alpha$, and red = [S~{\sc ii}]
$\lambda\lambda$6717,6731 (from Smith et al.\ 2000).  (b) Color image
from MSX data, with blue = Band A, green = Band C, and red = Band E.
(c) Color image from IRAS data, with blue = 25 $\mu$m, green = 60
$\mu$m, and red = 100 $\mu$m.  (d) False color image of the 3.4 cm
radio continuum measured by the Parkes telescope (see Huchtmeier \&
Day 1975).  The dashed rectangle in Panel (c) is the region over which
we integrated the total fluxes for the nebula at all wavelengths.
Galactic coordinates are measured in degrees.  Eta Carinae and the
center of Tr16 are located at $l,b$=287\arcdeg36\arcmin,
$-$0\arcdeg38\arcmin, while Tr14 is at $l,b$=287\arcdeg25\arcmin,
$-$0\arcdeg36\arcmin.}
\end{figure*}

\subsection{Radio Data} 

In order to measure the total radio continuum flux as a proxy for the
ionizing flux in the nebula, we need a radio continuum map that covers
the whole Carina region with single-dish data.  This is because flux
at low spatial frequencies is lost in interferometric data like that
from the Molonglo Observatory Synthesis Telescope (MOST) presented by
Whiteoak (1994), although we will use those data for investigating the
multiwavelength morphology.

To date, the best single-dish radio continuum map of Carina is still
the 3.4cm continuum map made over 30 years ago by Huchtmeier \& Day
(1975).  This map, obtained with the 64m Parkes telescope, has an
effective spatial resolution of $\sim$2$\farcm$6, comparable to the
{\it IRAS} data.  Unfortunately, the data were not available to us in
digital form.  Therefore, we used a crude technique to digitize the
contour map published by Huchtmeier \& Day (1975) into a FITS file
(i.e. we drew a grid over the original contour map, and read-off the
intensity in each pixel, interpolating when necessary).  This yielded
a map of the 3.4cm intensity over most of the nebula, shown in Figure
1$d$, which can then be used to measure the integrated 3.4cm flux and
the spatial distribution of free-free radio emission.  This map is
missing some low surface brightness emission in the outer nebula,
which can be seen in the H$\alpha$ image but is below the lowest radio
contour level.  This low-level emission should contribute less than
$\sim$10\% of the total flux, which is less than the uncertainty.

Additional radio data will be considered later in this paper as well.
To examine the spatial distribution of 0.843 GHz radio continuum
emission at higher spatial resolution (roughly 30\arcsec), we will
examine the MOST maps from Whiteoak (1994), as mentioned above.  These
data were obtained from the online Molonglo Galactic Plane Survey
(MGPS) made with the MOST telescope.\footnote{\tt
www.physics.usyd.edu.au/astrop/most.}  To compare the distribution of
molecular gas in giant molecular clouds, we will use the emission maps
of Carina obtained by the recent $^{12}$CO(1--0) NANTEN survey of
Yonekura et al.\ (2005), which have an effective beam size of
2$\farcm$7.

\section{INTEGRATED NEBULAR PROPERTIES}

To evaluate the global mass and energy budget of Carina, we wish to
measure the total IR luminosity, the total free-free radio continuum
emission, and the total H$\alpha$ flux emitted by the entire nebula.
We begin with thermal dust emission.

\subsection{The SED and the Total IR Luminosity}

We regard the presence of the exceedingly bright object $\eta$~Carinae
as anomalous compared to most giant H~{\sc ii} regions, since
$\eta$~Car is in a very brief eruptive evolutionary phase.  Therefore,
we subtracted its direct influence from the total integrated fluxes.
Since nearly all of its stellar radiation is reprocessed by
circumstellar dust and escapes in the IR (Smith et al.\ 2003b),
radiation from $\eta$~Car no longer contributes to the radiative
energy budget of the H~{\sc ii} region. This is appropriate, since in
many of the images $\eta$~Car itself is saturated, and thus, the
global integrated flux including $\eta$ Car would be incorrect
anyway. Before summing the total observed flux for each filter within
the box in Figure 1$c$, we used {\sc imedit} in {\sc iraf} to
carefully interpolate over $\eta$~Carinae, removing its flux from each
image.  This process was admittedly subjective because of possible
confusion between $\eta$~Car and the diffuse source Car~{\sc ii} (the
Keyhole Nebula) at some wavelengths.  However, the integrated IR to
radio spectral energy distribution (SED) of $\eta$~Car is known well
(e.g., Cox et al.\ 1995; Morris et al.\ 1999; Smith et al.\ 2003b;
Brooks et al.\ 2005), providing an independent check on our results,
and the complex structure of the Keyhole region including $\eta$~Car
has been observed at higher resolution at optical to radio wavelengths
(Brooks et al.\ 1998, 2000, 2001, 2005; Cox \& Bronfmann 1995; Smith
2002b; Whiteoak 1994), which helps minimize confusion.  This
subtraction is important in H$\alpha$ and in the mid-IR at 8--30
$\micron$, where $\eta$~Car itself contributes nearly half of the
total flux from the Carina Nebula.  However, at the more critical
far-IR and radio wavelengths of interest here, $\eta$~Car's
contribution is minimal anyway, as it provides less than 3\% and 0.5\%
of the total emission at 60 and 100 $\micron$, respectively.

\begin{table}\begin{minipage}{3.4in}
\caption{Total Measured Fluxes}\scriptsize
\begin{tabular}{@{}lcccc}\hline\hline
$\lambda$($\micron$) &I.D.  &Units  &Flux &Error\\ \hline
0.501	&[O~{\sc iii}]	&erg s$^{-1}$ cm$^{-2}$	&2.12(-7)	&5\%	\\
0.656	&H$\alpha$	&erg s$^{-1}$ cm$^{-2}$	&6.40(-7)	&5\%	\\
8.3	&MSX A  	&Jy	&1.54(4)	&10\%	\\
12.1	&MSX C  	&Jy	&4.26(4)	&10\%	\\
14.7	&MSX D  	&Jy	&3.49(4)	&10\%	\\
21.3	&MSX E  	&Jy	&9.55(4)	&10\%	\\
12	&IRAS 1 	&Jy	&4.26(4)	&10\%	\\
25	&IRAS 2 	&Jy	&1.53(5)	&10\%	\\
60	&IRAS 3 	&Jy	&8.75(5)	&10\%	\\
100	&IRAS 4 	&Jy	&1.43(6)	&10\%	\\
34000	&Parkes 	&Jy	&1.43(3)	&20\%	\\ \hline
\hline
\end{tabular}
\end{minipage}
\end{table}

Table 1 lists integrated fluxes at each wavelength for the entire
Carina Nebula, summed over the region within the dashed rectangular
box in Figure 1$c$.  This includes most of the observed filaments
associated with Carina, but attempts to exclude the emission from the
nearby H~{\sc ii} region NGC3324.  The total fluxes have been sky
subtracted by sampling a region of blank sky at the bottom of the
images, but subtraction of the irregularly varying diffuse emission
from the Galactic plane itself was problematic. This was the main
source of uncertainty quoted in the last column of Table 1.  By
sampling nearby regions in the Galactic plane immediately adjacent to
Carina, we found that this diffuse emission may contribute as much as
$\sim$5\% at the shorter mid-IR wavelengths, and possibly as much as
10\% of the total flux at 60 and 100 $\micron$.  The fluxes in Table 1
have therefore been reduced by these fractions.\footnote{The high
precision to which the 12.1 $\mu$m MSX flux matches the 12 $\mu$m IRAS
flux is fortuitous.  However, even general agreement despite the
differences in resolution and sensitivity reassures us that our method
of measuring the integrated flux is sound.}  The H$\alpha$ and [O~{\sc
iii}] images were background subtracted before taking the total flux
by fitting a surface polynomial to the background Galactic plane
emission.  These two images were also continuum subtracted before
measuring the total flux using a scaled version of the R band image
for H$\alpha$ and an average of the R and B band images for [O~{\sc
iii}].  Continuum subtraction was not perfect, but the difference in
total line flux before and after continuum subtraction was $<$3\% in
both filters.

The observed total fluxes from Carina are plotted in Figure 2, which
shows the IR-to-radio SED of the whole nebula.  It is striking how
similar the SED of Carina is to that of a typical ULIRG (e.g., Sanders
\& Mirabel 1996).  A fit to this SED is shown by the solid curve,
using emission from three optically thin greybody (i.e. emissivity
$\propto\lambda^{-1}$) components at T=220, 80, and 35 K, plus a
simple component from optically thin thermal bremsstrahlung radio
continuum emission.  We chose 3 specific dust temperatures because
this is the minimum number that can account for the observed SED's
shape, but it is possible that the spectrum at mid-IR wavelengths can
be approximated by a range of temperatures.  However, decomposing the
SED into additional components would not significantly alter the
estimate of the total luminosity and dust mass (Table 2).  Fluxes in
the 8.3, 12--12.1, and 21.3 $\micron$ filters are allowed to be
somewhat above this fit because they may contain strong emission from
PAH and silicate emission features at 8.6, 9.7, 11.3, and 22
$\micron$.  These emission features suggest that the 220 K component
in Figure 2 and Table 2 does not truly represent continuum emission
from hot grains.  Hence, the 220 K component's dust mass in Table 2 is
listed in parentheses.  In any case, the shorter IR wavelengths are
unimportant in the mass budget.

\begin{figure*}\begin{center}
\epsfig{file=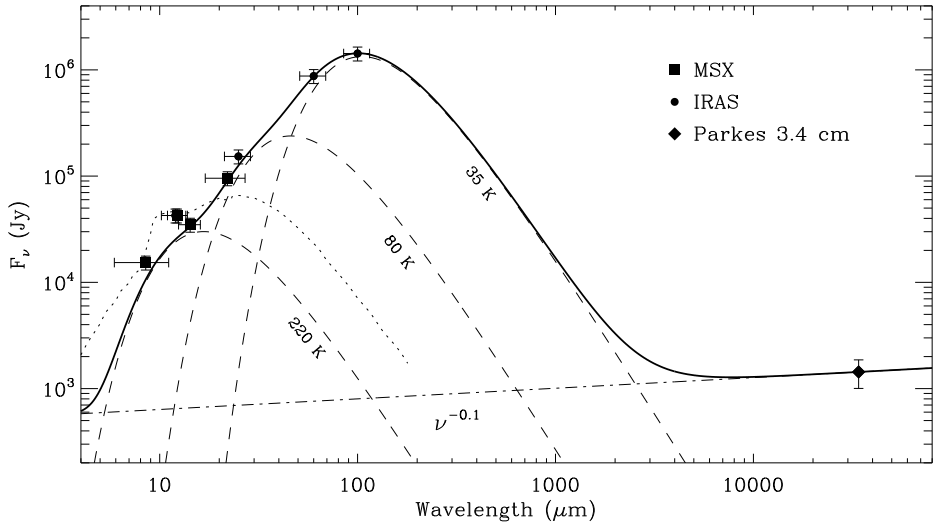,width=5.0in}
\end{center}
\caption{Global IR-to-radio spectral energy distribution of the Carina
  Nebula.  Fluxes were measured within the boundary of the large
  dashed rectangle in Fig.\ 1$c$, and all have had the flux from
  $\eta$ Car subtracted (the IR spectrum of $\eta$ Car measured by ISO
  is shown by a dotted curve; see Smith et al.\ 2003b; Morris et al.\
  1999).  Sky and background subtraction were performed as noted in
  the text.}
\end{figure*}

\begin{table}\begin{minipage}{3.4in}
\caption{IR Dust Luminosity and Mass}\scriptsize
\begin{tabular}{@{}lcc}\hline\hline
T(K) &L/L$_{\odot}$  &M$_{d}$/M$_{\odot}$ \\ \hline
35	&7.6$\times$10$^6$	&9590	\\
80	&3.1$\times$10$^6$	&28			\\
220	&1.1$\times$10$^6$	&(0.23)			\\
Total	&1.18$\times$10$^7$	&9620	\\ \hline
\hline
\end{tabular}
\end{minipage}
\end{table}

The luminosities for each of these three component fits and the total
are listed in Table 2.  As long as the dust grains are small ($a<$0.2
$\micron$), the dust mass required to emit this IR luminosity can be
expressed independently of the grain radius and emissivity (see Smith
\& Gehrz 2005), so that

\begin{equation}
M_{\rm dust}=[(100\rho)/(3\sigma T^6)] \times L_{\rm IR}
\end{equation}

\noindent where $\sigma$ is the Stefan-Boltzmann constant.  This
relation was used to derive the dust masses for each component in
Table 2, assuming a typical grain density of about 3 g cm$^{-3}$.
Given our assumption of optically thin emission, the true dust mass
may be higher than that given in Table 2 if additional mass can be
hidden in dense cloud cores that are optically thick even at far-IR
wavelengths.  However, given that the peak value of the 60 $\micron$
optical depth is only 0.005 and the average is much lower than that
(see \S4.1 and Fig.\ 4b), any optically thick clumps must constitute a
tiny fraction of the solid angle of the nebula, so that the bulk of
the far-IR emission is, in fact, dominated by optically thin emitting
dust.  Optically thick cores may cause submillimeter or millimeter
emission in excess of the 35 K component in Figure 2, and future
observations will be useful in that regard.  However, preliminary
analysis of a 1mm survey of the brightest parts of the nebula (K.\
Brooks et al.\ in prep.; see Brooks et al.\ 2005), with a 2$\sigma$
sensitivity equivalent to a 10 M$_{\odot}$ core at 10 K, suggest an
accumulated gas mass in such cold cores of no more than 5,000
M$_{\odot}$.  Similarly, Yonekura et al.\ (2005) detected 15 cores in
C$^{18}$O emission, with a combined mass of only 2.6$\times$10$^3$
M$_{\odot}$.  Also, the near-perfect correlation between the 60
$\micron$ optical depth and the 8.6 $\micron$ PAH emission (see \S
4.1) argues that optically thick cores contribute little to the global
60-100 $\micron$ emission from Carina.

The total integrated IR luminosity from these three dust components
amounts to $\sim$1.2$\times$10$^7$ L$_{\odot}$ at a distance of 2.3
kpc.  Thus, dust absorbs and reprocesses 50--60\% of the available
bolometric luminosity from all stars in Carina, which is
$\sim$2$\times$10$^7$ L$_{\odot}$ (Paper I).\footnote{Excluding
$\eta$~Car's luminosity of 5$\times$10$^6$ L$_{\odot}$ (Smith et al.\
2003b).  Most of $\eta$ Car's radiation escapes in the mid-IR, which
has already been subtracted directly from our data.}

The coolest 35 K component, probably dominated by heating from FUV
photons, contributes about 65\% of the IR luminosity or about 40\% of
the total luminosity.  This suggests that about 20\% of the FUV
radiation escapes through holes in the nebula or is processed by PAH
emission, if $L_{FUV}/L_{Bol}\simeq$0.5 (see Paper I).  The 35 K
component dominates the emission seen at far-IR wavelengths, in
agreement with earlier studies of the far-IR emission from smaller
sections of the nebula that inferred dust temperatures of 30--40 K
(Harvey et al.\ 1979; Ghosh et al.\ 1988).

The warmer 220 and 80 K components that dominate at shorter
wavelengths absorb and re-emit the remaining 35\% of the IR
luminosity, corresponding to only $\sim$20\% of the total stellar
luminosity.  As noted earlier, the 220 K population may be erroneous,
as would be its associated dust mass estimate, since the shorter
mid-IR wavelengths may be dominated by diffuse PAH emission.  However,
the 80 K component probably represents warm grains surviving in the
interior of the H~{\sc ii} region.  Throughout most of the nebula,
they are heated stochastically by Ly$\alpha$ photons, and therefore,
ultimately by Lyman continuum photons (see \S 4.1).  Close to the O
stars, direct heating of dust by stellar UV continuum radiation may
also be important.  These 20-25 $\mu$m emitting grains are often seen
as filamentary structures seen in the MSX band-E image (Fig.\ 1b);
they may reside in dense swept-up shells where the O star winds
confront the phoevaporative flows from ionization fronts.

\subsection{The Mass Budget}

The cool 35 K component dominates the mass of emitting dust in the
nebula.  While the warmer components make a substantial contribution
to the luminosity, they are irrelevant to the total dust mass.  The
total mass revealed by thermal dust emission is about 9600 M$_{\odot}$
(Table 2).  The uncertainty in this dust mass is large, perhaps of
order $\pm$20\% because of potential errors in background subtraction
and uncertain grain properties.\footnote{The mass estimate is
dominated by the 35 K component in Fig.\ 2).  This component can't be
hotter or more luminous, but there could be additional mass hidden in
cool grains that we are not sensitive to because we do not yet have a
global estimate of the submm luminosity of the nebula, as noted
earlier.}

The total mass of molecular gas in the Carina region is estimated to
be 6.7$\times$10$^5$ M$_{\odot}$ from the Columbia CO survey of the
Galactic plane (Grabelsky et al.\ 1988).  The relevant molecular mass
is even less than this, since this value was estimated over a region
larger than the box in Figure 1$c$.  Yonekura et al.\ (2005) give a
smaller value of 3.5$\times$10$^5$ M$_{\odot}$.  As explained below in
\S 4, we favor an interpretation where the mass traced by the far-IR
dust emission is probably mixed-in with warm atomic gas in PDRs behind
ionization fronts because it has a different spatial distribution than
the CO clouds (Fig.\ 3).  Thus, the gas:dust mass ratio inferred from
comparing molecular gas and dust is erroneous because the warm
emitting dust and the CO gas trace different components of the nebula.

Instead, if we assume that the far-IR emission from $\sim$35~K dust
traces warm atomic gas in PDRs, then a typical gas:dust mass ratio of
100 there would indicate a total gas mass of $\sim$10$^6$ M$_{\odot}$.
This, in turn, would indicate that the warm gas in PDRs dominates over
molecular clouds in the total mass budget of the nebula.  Brooks et
al.\ (2003), however, find that most of the gas-phase carbon is in CO,
so a large reservoir of additional mass may be present that is not
traced by the warm $\sim$35 K dust.  The gas currently in the
molecular phase is less than about one-third of the total mass budget.
Thus, adding the molecular and PDR mass together, about
1-2$\times$10$^6$ M$_{\odot}$ in the nebula may still be available for
further star formation, providing that it can be swept up into dense
enough clouds (see \S 4.4).

What is the mass ratio between this gas and the stars in Carina?  A
complete stellar census is not yet available, but perhaps we can draw
preliminary conclusions by scaling from a well studied region like the
Orion Nebula Cluster (ONC).  Hillenbrand \& Hartman (1998), for
example, find a total number of stars of 2260 and a total stellar mass
of 1800 M$_{\odot}$ for members of the ONC.  Now, the number of stars
in the ONC with masses above about 10 M$_{\odot}$ is 6 and the number
of O-type stars is 2, while for the first-generation exposed clusters
in Carina the corresponding numbers are 127 and 70 (Paper I).  This is
certainly not a complete census for Carina since there are many early
B stars not included in Paper I, and some O stars may still be
obscured (e.g., Sanchawala et al.\ 2007).  Scaling from Orion, then,
we find likely values for the total number of first-generation stars
in Carina of about 5-8$\times$10$^4$, harboring a total stellar mass
of 4-6$\times$10$^4$ M$_{\odot}$.  Errors may be as much as a factor
of 2 with this crude method, but are likely to be an underestimate if
wrong.  We undertook a more detailed analysis where we scaled the
Trapezium mass function of Muench et al.\ (2002) so that the slope at
high masses agreed with a mass function from the spectral types in
Paper I, and then integtrated over a range of 0.08-150 M$_{\odot}$ to
get the total mass, but this produced similar results with similar
uncertainty.  In any case, the total stellar mass in the first
generation stars in Carina is securely above 2$\times$10$^4$
M$_{\odot}$, placing it among the most massive stellar groupings in
the Galaxy.  This is in agreement with its large Lyman continuum
luminosity.

It would appear from this value for the total stellar mass that the
global star formation efficiency in Carina has been less than 10\%.
However, two important considerations suggest that this would likely
be a severe underestimate, and that the star formation efficiency of
the first generation clusters could be as high as 30-50\%.  First, the
total gas mass is derived for a very large region, and likely contains
mostly pristine gas that did not participate in the formation of the
first generation star clusters like Tr14 and 16.  The mass of gas
which has been expelled from the inner region of the nebula that was
formerly the molecular cloud core is unclear, but it is probably a
small fraction of the total nebular mass.  Second, the stellar census
is surely incomplete, because it ignores the vast number of second
generation stars that are now forming or have recently formed around
the periphery of the nebula, which may be comparable to the first
generation.  An ongoing analysis of Spitzer data (Smith et al., in
prep.) suggests that the South Pillar region alone contains well over
10$^4$ new stars.  Thus, a census of this younger second generation
will be quite interesting.

\subsection{The Kinetic Energy Budget, Diffuse X-rays, and a Previous Supernova?}

If we take the mass of warm atomic gas inferred in the previous
section at face value as the dominant mass component, we can evaluate
the total kinetic energy budget of the nebula.  We found a total gas
mass of $\sim$10$^6$ M$_{\odot}$ from the cool dust mass for an
assumed normal gas:dust mass ratio of 100.

High-resolution spectroscopic observations of ionized gas in Carina
over the past decades have repeatedly established a ubiquitous line
splitting of 35--40 km s$^{-1}$ across the entire region (e.g.,
Meaburn et al.\ 1984; Azcarate et al.\ 1981; Deharveng \& Maucherat
1975; Walborn \& Hesser 1975; Smith et al.\ 2004a).  If we then take
$\pm$20 km s$^{-1}$ as representative of the bulk expansion velocity
of the developing superbubble cavity (Smith et al.\ 2000), the implied
total kinetic energy of the nebula would be of order
8$\times$10$^{51}$ ergs.  This agrees well with the value of
$\sim$9$\times$10$^{51}$ ergs estimated on different grounds by Smith
et al.\ (2000); it was the energy needed to sweep up the cavity to the
observed size assuming a homogeneous ambient medium.


Although the large kinetic energy budget of the nebula allows for
input from one or more past supernovae, there is no {\it need} to
invoke a previous supernova to account for it.  The present kinetic
energy of the nebula we derive here is only 30\% of the available
kinetic energy of 2.6$\times$10$^{52}$ ergs supplied by stellar winds
throughout the $\sim$3 Myr lifetime of the nebula (Paper I).  Thus,
the observed kinetic energy budget gives no evidence to alter the view
that the Carina Nebula is in an early phase of its expansion still
dominated by stellar winds and radiation pressure.\footnote{There is
one interesting caveat here, however. Recent studies of UV spectra of
O stars (e.g., Fullerton et al.\ 2006; Bouret et al.\ 2005) suggest
that their winds are highly clumped and that their mass-loss rates may
be lower than the rates adopted in Paper I, which were from estimates
based on moderate clumping factors (Repolust et al.\ 2004).  If the
mass-loss rates are indeed reduced much further than assumed in Paper
I, then the input mechanical energy from stellar winds may become
comparable to the observed expansion energy of the nebula.}  This is
supported by the lack of any evidence for significant non-thermal
radio continuum emission from the nebula.  It is also in agreement
with Brooks et al.\ (2003), who find that stellar winds alone can
account for the kinematics of the molecular-line data.  In that case,
the diffuse soft X-ray emission in Carina (Seward et al.\ 1979; Seward
\& Chlebowski 1982) could arise from a shock between stellar winds
expanding from O stars in Tr14 and Tr16 as they collide with the
evaporative PDR flow from the surrounding molecular clouds.  The
intricate structure of the evaporating pillars would make the geometry
of this shocked material very irregular, causing strong variations
along adjacent lines of sight, while still being present across the
whole region.  Thus, it is plausible that the high-velocity absorption
components (e.g., Walborn et al.\ 2007, 1984; Walborn \& Hesser 1975;
Danks et al.\ 2001) may result within these wind interaction regions
as well.

\subsection{Tracers of the Ionized Gas}

The total flux of free-free radio continuum emission is often taken as
a diagnostic of the total Lyman continuum luminosity of an H~{\sc ii}
region, because it avoids potentially large uncertainties in
extinction.  The total number of H-ionizing photons absorbed by the
gas can be expressed as

\begin{equation}
Q_H = 9.09\times10^{46} \ T_4^{-0.45} 
  \big{(}\frac{\nu}{8.82 GHz}\big{)}^{0.1} S_{\nu} D^2 \ s^{-1}
\end{equation}

\noindent where $T_4$ is the assumed electron temperature in units of
10$^4$ K, $S_{\nu}$ is the observed flux density in Jy at frequencey
$\nu$, and $D$ is the distance in kpc.  In equation (2), we have
assumed that the n(He)/n(H) abundance is 0.1, that the volume of the
He$^+$ region is half of the H$^+$ region, and that the fraction of He
recombination photons that can ionize H is $f_i$=0.65 (e.g., Simpson
\& Rubin 1990).  In that case, the factor
(1+$f_i$$<$He$^+$/[H$^+$+He$^+$]$>$)$^{-1}$ is 0.969; it would be
0.939 with equal volumes for the He$^+$ and H$^+$ zones, changing our
value for Q$_H$ by an insignificant 3\%.

From Table 1, we measure a total 3.4 cm (8.82 GHz) flux of
$S_{\nu}$=1430 Jy for the whole nebula, integrated over the large
rectangular area in Figure 1$c$.  Assuming T$_e$=10$^4$ K in equation
2, this flux translates to a total Lyman continuum photon luminosity
of $Q_H\simeq$6.9$\times$10$^{50}$ s$^{-1}$.  This can be compared to
our estimate in Paper I of the total Lyman continuum output of the
known OB stars in Carina of $Q_H$=9.1$\times$10$^{50}$ s$^{-1}$.
Thus, it would appear from this analysis that roughly 25\% of the
ionizing photons may be able to leak out of Carina.\footnote{Note,
however, that this is the observed value only at the present time,
when the dusty Homunculus nebula and $\eta$ Car's dense wind quench
the UV output of $\eta$ Car and its putative companion (see Paper I).
In the recent past (before 1843), the total ionizing flux output of
the stars in Carina may have been as high as
$Q_H$=11.5$\times$10$^{50}$ s$^{-1}$ (Paper I).  The ionizing output
in this previous state may be relevant even to present observations,
since the recombination timescale in the outer parts of the nebula
would be about 50 yr for a typical electron density at ionization
fronts of a few $\times$10$^{3}$ cm$^{-3}$ (Smith et al.\ 2004b),
while the light travel time to some outer parts of the nebula is
comparable.}  This escape fraction will need to be larger if there
remains a significant number of hot stars in Carina that are obscured
at visual wavelengths and were not included in the census in Paper~I.

Also using radio continuum observations, Brooks et al.\ (2001)
estimated the number of H-ionizing photons absorbed locally by the
radio sources Car~I and Car~II (associated with Tr14 and Tr16,
respectively).  Comparing these values to the expected Lyman continuum
output from each cluster in Paper I, it appears that Car~I absorbs
5.5\% of Tr14's ionizing radiation, while Car~II absorbs much less
than 1\% of the ionizing photons from Tr16.  This tells us that the
vast majority of ionizing radiation from these clusters escapes to
large distances in the nebula.  It also means that Tr14 is more
involved in dense gas than Tr16, providing another clue that it may be
somewhat younger than Tr16.


The total H$\alpha$ luminosity that escapes the nebula (corrected for
the average interstellar extinction but not local non-uniform
extinction from dark clouds within the nebula) is about 10$^5$
L$_{\odot}$, and the escaping luminosity in the [O~{\sc iii}]
$\lambda$5007 line is about 3$\times$10$^4$ L$_{\odot}$.  This
H$\alpha$ luminosity is about 1/3 that of 30 Doradus (Kennicutt 1984).

In studies of extragalactic H~{\sc ii} regions and star forming
galaxies, the total H$\alpha$ luminosity is also used to derive the
number of hydrogen ionizing photons.  Following Kennicutt (1998),

\begin{equation}
Q_H = 2.93\times10^{45} \ \big{(}\frac{L_{H\alpha}}{L_{\odot}}\big{)} \ s^{-1}
\end{equation}

\noindent where L$_{H\alpha}$ is the luminosity in the H$\alpha$
emission line.  For our measured value of L$_{H\alpha}\simeq$10$^5$
L$_{\odot}$ for the total H$\alpha$ line luminosity in Carina, we then
have $Q_H\simeq$3$\times$10$^{50}$ s$^{-1}$.  Thus, with standard
assumptions the observed H$\alpha$ flux appears to significanty
underestimate the total Lyman continuum flux as measured from the
radio continuum (about 40--50\%), and it severely (factor of 3)
underestimates the expected Lyman continuum luminosity from known O
stars (Paper I). Perhaps this is because the H$\alpha$ line suffers
from severe non-uniform extinction.  The required average extinction
would be an additional $A_R\simeq$0.9 mag.  We corrected the H$\alpha$
flux for the small amount of average interstellar extinction toward
Carina, but not for local extinction; we will see below in \S 4.2 that
non-uniform local extinction is severe.

The ratio of the global [O~{\sc iii}] $\lambda$5007 luminosity to that
of H$\alpha$ for the whole nebula is about 0.3.  The observed and
dereddened ratio of [O~{\sc iii}] $\lambda$5007 to H$\alpha$ in the
brightest inner parts of the nebula is $\sim$1 (Smith et al.\ 2004b).
This means that [O~{\sc iii}] is more centrally concentrated than
H$\alpha$, and again, may be more heavily absorbed there due to
non-uniform extinction, which is in fact the case as we discuss later
in \S 4.2.  This suggests caution when interpreting [O~{\sc
iii}]/H$\alpha$ ratios in extragalactic H~{\sc ii} regions.  As noted
in Paper I, the large ionizing photon luminosity places Carina among
the most extreme star forming regions in our Galaxy, such as W49 and
NGC3603, although not quite as extreme as the Arches cluster in the
Galactic center or 30 Dor.

\begin{figure*}\begin{center}
\epsfig{file=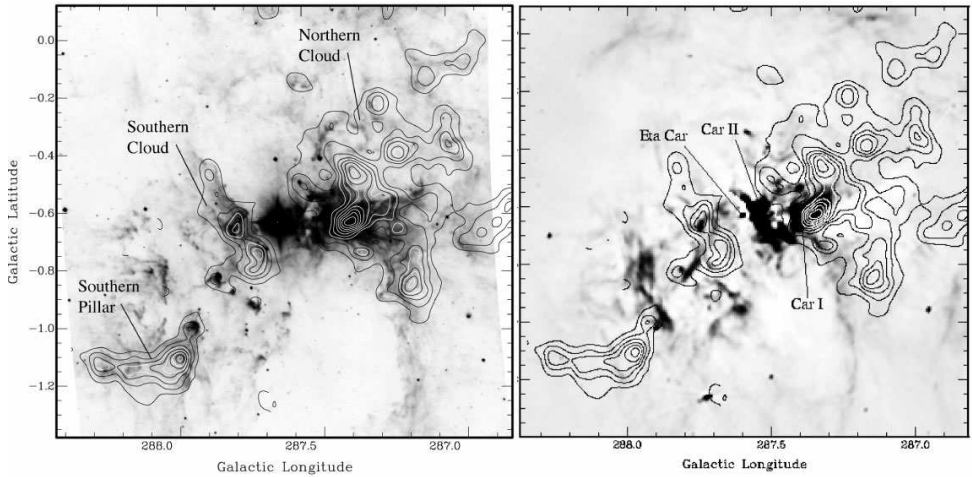,width=5.1in}
\end{center}
\caption{Molecular gas distribution in the Carina Nebula, with
contours from the NANTEN $^{12}$CO(1--0) survey of Yonekura et al.\
2005.  These contours are superposed on images at 8~$\mu$m from MSX
(left; Smith et al.\ 2000) and in the 0.843 GHz continuum from MOST
(right; Whiteoak 1994).}
\end{figure*}

\begin{figure*}\begin{center}
\epsfig{file=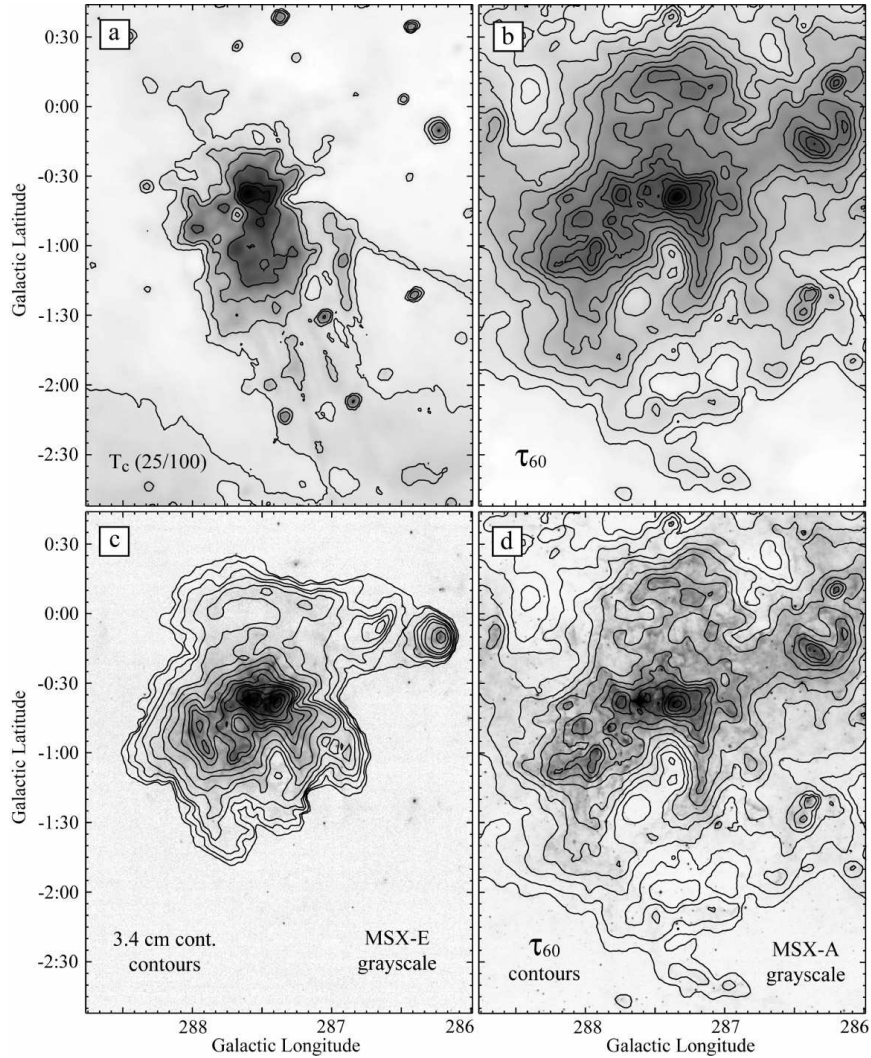,width=4.5in}
\end{center}
\caption{(a) Far-IR dust color temperature made from a ratio of the
  IRAS 25 and 100 $\mu$m maps.  Contours are drawn at 52, 54, 56, 60,
  65, and 70 K.  (b) 60 $\mu$m emitting optical depth from the IRAS 60
  $\mu$m image and the 25/100 $\mu$m dust color temperature map in
  Panel a.  Optical depth contours are drawn at 3, 4, 6, 9, 13, 20,
  30, 50, 70, 100, 140, 200, 320, and 500 $\times$10$^{-5}$, (c) Radio
  contours from Fig.\ 1d over the MSX band-E (21.3 $\mu$m) map from
  Figure 1$b$.  (d) 60 $\mu$m optical depth contours over the MSX
  Band-A (8.3 $\mu$m) image.}
\end{figure*}

\section{LARGE-SCALE MULTIWAVELENGTH MORPHOLOGY}

\subsection{Dust Distrubution}

Figures 4$a$ and 4$b$ show the far-IR dust color temperature and 60
$\mu$m emitting optical depth, respectively.  The temperature map is
made from a ratio of the 25 and 100 $\mu$m IRAS images, while the
optical depth map uses this temperature map and the 60 $\mu$m IRAS
image.  As noted earlier, we did not use deconvolved images with
enhanced resolution, because such images can lead to erroneous
features in ratio images, and we are interested primarily in the
large-scale distribution (note that we smoothed the 25 $\mu$m map to
match the spatial resolution of the longer wavelength data).  For each
pixel, the dust color temperature and optical depth were calculated in
the usual way from ratio images (e.g., Smith et al.\ 2003b).  However,
we caution that the numerical values of T and $\tau$ in the maps are
not necessarily the actual temperature or optical depth of grains in
the nebula, since each line of sight may contain multiple grain
populations at different temperatures.  For example, the SED in Figure
2 shows that the 25 $\mu$m flux is dominated by warm $\sim$80 K grains
across much of the nebula, whereas the 100 $\mu$m flux is dominated by
cooler $\sim$35~K dust.  Therefore, contour levels in Fig.\ 4 are
average values, but are still useful in assessing trends in {\it
relative} dust temperature and column density.


We find an excellent spatial correlation between the large-scale
distribution of radio continuum and the 21.3 $\mu$m Band E emission
from MSX, shown in Figure 4c.  This was noted on smaller scales by
Rathborne et al.\ (2004) and Cox (2005) as well.  This is consistent
with the interpretation that warm grains in the interior of H~{\sc ii}
regions are intermixed with and heated {\it in situ} by emission from
the ionized gas itself.  For the case B recombination that prevails in
most H~{\sc ii} regions, Ly$\alpha$ photons trapped in the nebula
resonantly scatter many times before being absorbed by dust.  Spitzer
(1978) showed that for graphite grains bathed in ionized gas with $n_H
\simeq$10$^3$ cm$^{-3}$ and roughly Solar metallicity, the equilibrium
temperature from the balance of absorption of trapped Ly$\alpha$
photons and IR emission would be about 80~K.  This is, coincidentally,
the same temperature we derive for the warm component that dominates
the SED at 20--30 $\mu$m (Fig.\ 2).  By comparison, a simple estimate
indicates that grain-gas collisional energy gain or direct heating by
stellar photons is less by at least two orders of magnitude over most
of the nebula, although direct UV continuum heating may become
important close to the O stars.  Some grains in Carina show IR
emission features similar to SNRs like Cas A (Chan \& Onaka 2000), but
there appears to be no need to invoke grain heating by a recent
supernova to explain the luminosity or temperature of the warm grains.

In general, these results indicate that diffuse 20--30 $\mu$m emission
that occupies the interiors of similar shell structures can be taken
as an excellent tracer of dust that is mixed with dense ionized gas in
H~{\sc ii} regions.  The association of warm grains with ionized gas
explains the red emission in the interior of the nebula in the color
MSX image in Figure 1$b$, while PAH emission from the surfaces of
molecular clouds is seen in blue/green at larger distances from the
center.  This pattern clearly holds true in the adjacent region
NGC3324 as well (Fig.\ 1b), as does the good correlation between 20-25
$\mu$m emission and radio continuum (Fig.\ 4c). This interpretation is
probably applicable to H~{\sc ii} regions in general.  Such features
are common in 21.3 $\mu$m MSX data or in 24 $\mu$m images taken by
Spitzer throughout the Galactic plane (e.g., Churchwell et al.\ 2006).
It would be interesting to see if this type of feature is less obvious
at lower metallicity, where the dust:gas ratio is lower.


As interesting, perhaps, is the near-perfect correlation over large
scales between the 60~$\mu$m optical depth and PAH emission
shown in Figure 4$d$, ignoring differences in spatial resolution
between IRAS and MSX images.  The only instances where this
correlation is not obeyed are for point sources like $\eta$ Carinae,
embedded protostars, and cool giants in the field.  However, the
8.6~$\micron$ MSX Band-A emission from these unresolved sources is
primarily due to continuum emission from warm circumstellar grains
rather than PAH emission.  This provides spatially-resolved
confirmation that the cool 30--40 K dust that dominates the far-IR
luminosity measured by IRAS -- tracing the majority of the nebular
mass -- resides primarily in warm atomic zones in PDRs heated by the
FUV radiation field, and not in colder molecular cloud cores.  (The
100 $\mu$m optical depth map is essentially the same as the 60 $\mu$m
optical depth.)  The PAH and warm dust distributon is significantly
different from that of the molecular cloud cores when examined at the
level of detail seen in Figure 3.  On larger size scales where the
structure of PDRs is unresolved, The molecular gas and PAH+dust
emission do trace one another because the PAH mission comes from the
surfaces of molecular clouds.  This similar structure on large sizes
is due to the clumpy distribution of the gas, allowing FUV radiation
to penetrate to large radii in between the clumps.  This view is
compatible with the large scale distribution of high density molecular
gas seen in CO(4$-$3) and PDR emission seen in [C~{\sc i}] at 610
$\micron$ (Zhang et al.\ 2001).

Earlier we estimated that the difference between IR luminosity and the
inferred stellar FUV luminosity in Paper~I means that about 20\% of
this FUV luminosity escapes the nebula.  From the large-scale
distribution of dust in Figure 4$b$, this may be largely an effect of
the asymmetric and clumpy geometry.  There is a large hole toward
southern Galactic latitudes where a significant fraction of the FUV
radiation may escape.  It is even apparent in Fig.\ 4$a$ that higher
dust temperatures protrude in that direction.


\begin{figure*}\begin{center}
\epsfig{file=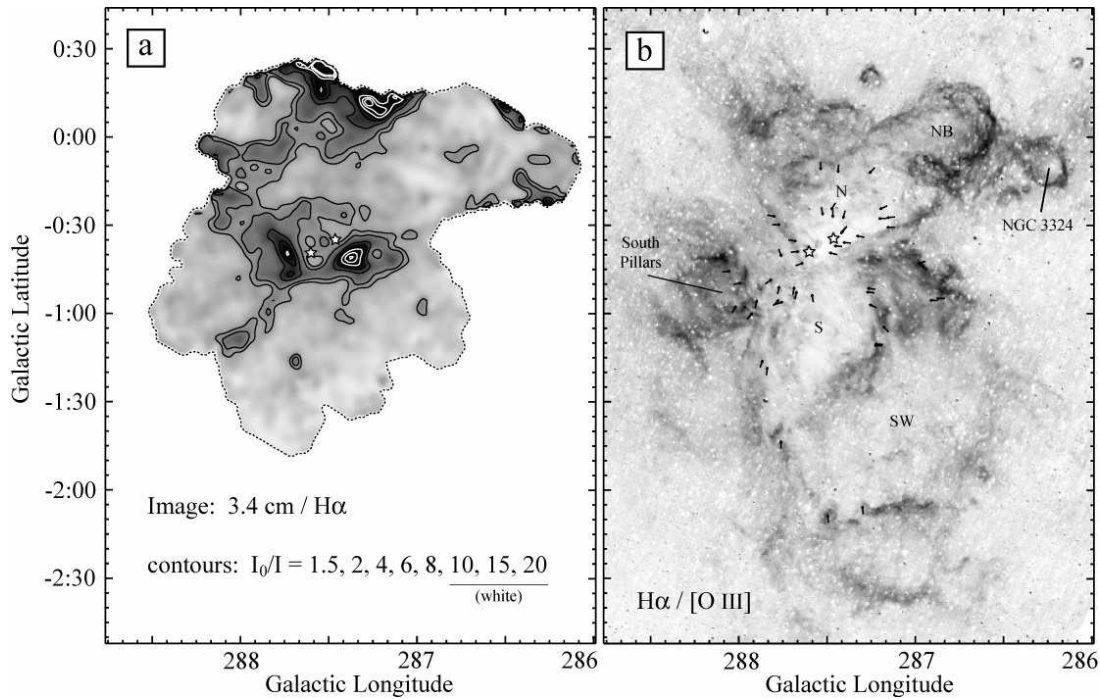,width=5.7in}
\end{center}
\caption{(a) Flux ratio image of 3.4 cm continuum to H$\alpha$.  The
  H$\alpha$ image was blurred to match the spatial resolution of the
  radio image in Fig.\ 1$d$.  Contours of $I_0/I$ are shown, where
  $I_0/I$ is derived with the radio/H$\alpha$ ratio normalized to
  unity across optically bright regions of the nebula (the dashed line
  is not a contour; it simply marks the boundary of the radio
  continuum image). $I_0/I$ therefore traces strong local variations
  in extinction across the nebula, where darker regions (higher values
  of $I_0/I$) have stronger line-of-sight extinction.  The regions of
  high extinction at the top edge of the nebula correspond to regions
  of low surface brightness. (b) Flux ratio image of H$\alpha$ to
  [O~{\sc iii}] $\lambda$5007, which shows shell-like structures that
  arise at the edges of large cavities.  Short arrows mark the
  orientations of ``directional indicators'' like cometary clouds and
  dust pillars (see text).  The two stars in both panels mark the
  approximate locations of the star clusters Tr14 and Tr16; Tr16 is
  the one on the left, although the stars in Tr16 are spread over a
  larger area than this symbol.}
\end{figure*}

\subsection{Extinction Distribution and 3D Geometry}

Figure 5$a$ shows an image of local variations of extinction across
the Carina Nebula, derived from a flux ratio image of 3.4~cm radio
continuum (Fig.\ 1$d$) to a smoothed version of the H$\alpha$ image.
The radio free-free emission and the intrinsic H$\alpha$ emission
should have the same spatial distribution, so dark regions in the
grayscale image correspond to regions of higher visual-wavelength
extinction.  Thus, we take the radio image as representing an
intrinsic surface brightness map of the nebula, $I_0$, whereas
H$\alpha$ suffers significant extinction from dust at visual
wavelengths, causing the apparent surface brightness $I$ to be very
different.  When we normalize these maps to have the same value in the
brighter and less obscured regions of the nebula (where $I_0/I$ has a
minimum value, set to 1), regions where ionized gas suffers
substantial absorption from foreground clouds have larger values of
$I_0/I$.  In all cases, the values of $I_0/I$ shown in Figure 5$a$ may
be underestimates, because this map gives the extinction averaged over
the 2$\farcm$6 beam of the 3.4cm image, whereas higher resolution
optical images show complex small scale structure.  The high values of
$I_0/I$ near the top of Fig.\ 5$a$ may be caused by low emission
levels near edges of the field.

In the central regions of the nebula, the distribution of local
extinction follows the general spatial distribution of the most
prominent CO emission as seen in Figure 3, as well as that of the
bright 8 $\mu$m PAH emission at molecular cloud surfaces.  The main
concentrations of molecular gas in the Northern Cloud, the Southern
Cloud, and the Southern Pillar are all clearly seen as local maxima in
the extinction map in Figure 5$a$.  The Southern Pillar seems less
prominent in the extinction map than in the CO map, suggesting that
the bulk of the dust and molecular gas there may be behind much of the
ionized gas along the line of sight.  Some of the other pillars are
known to be on the far side of the nebula as well, based on their
line-of-sight extinction patterns or kinematic properties (Smith et
al.\ 2004a, 2005b).

Part of the V-shaped dark dust lane that crosses the center of the
nebula can be seen as a bridge between the Northern and Southern
clouds in the extinction map, whereas this bridge is not clearly seen
in the CO map.  This indicates that the middle parts of this dust lane
have fairly low column density compared to other parts of the clouds,
with $I_0/I\simeq$2--3, corresponding to 6563 \AA\ extinction of
0.75--1.2 mag.  However, they figure prominently in optical images
because they are on the near side of the nebula.

The strongest regions of extinction, corresponding to the Northern and
Southern clouds, have peak values for $I_0/I$ of roughly 10 and 20,
respectively, corresponding to extinctions at 6563 \AA\ of 2.5 and 3.3
magnitudes.  These regions suffering high extinction correspond to
areas of strong intrinsic emission.  This non-uniform extinction that
blocks some of the brightest H$\alpha$ emission in the center of the
nebula is the qualitative explanation for why Q$_H$ derived from the
total H$\alpha$ luminosity is smaller than Q$_H$ derived from the
radio continuum.  Namely, even though we corrected the total H$\alpha$
luminosity for interstellar extinction to Carina using average values
of E(B--V) derived from optical spectra (e.g., Smith et al.\ 2004b),
this method did not correct for the non-uniform extinction in the
nebula, which is severe in some places.  This is because the optical
spectra one uses to derive the average extinction are dominated by the
emission from bright and relatively unobscured regions.  This may
present a potential problem for interpreting observations of very
distant unresolved H~{\sc ii} regions.


\subsection{Large Scale Flows, Shocks, and Bubbles}

Figure 5$b$ shows an image of the flux ratio of H$\alpha$ to [O~{\sc
iii}] $\lambda$5007 across the entire Carina Nebula.  Ignoring obvious
artifacts from bright stars, this gives an excellent view of the
large-scale shell structures that define the boundaries of the nebula.
Aside from the South Pillar region (see Smith et al.\ 2000), we draw
attention to four main cavities/shells in Carina.  The large cavities
labeled N and S in Figure 5$b$, seen to the north and south of the
main Tr14 and Tr16 clusters, are the two halves of what appears to be
the main bipolar cavity of the nebula.  Additional shells are the
northern blister (NB) and the large south-west shell (SW).  There are
additional shells as well to the SW direction, as noted by Smith et
al.\ (2000).  These shells are reminiscent of the main structural
components of 30 Dor (e.g., Wang \& Helfand 1991).

OB stars in Carina are distributed around the region, unlike
concentrated super star clusters such as NGC3603 and the Arches
cluster in the Galactic center (Crowther \& Dessart 1998; Figer et
al.\ 1999).  However, we can gain some insight into which star
clusters blew which bubbles by examining the detailed morphology of
the nebula.  Much as the proplyds in Orion all point toward
$\theta^1$C Ori (e.g., Bally et al.\ 2000), Carina contains many dust
pillars and cometary clouds, like those discussed by Smith et al.\
(2003a; 2004b), pointing toward the source of ionization or stellar
winds that shaped them.  Using unambiguous head-tail structures and
well-defined dust pillars as tracers, we measured coordinates and
position angles for several ``directional indicators'' throughout the
Carina Nebula as seen in H$\alpha$ images from the same dataset used
by Smith et al.\ (2003a).  For each of these directional indicators,
we plot an arrow in Figure 5$b$ over the H$\alpha$/[O~{\sc iii}] image
of the nebula.  We find that arrows in the southern S shell, including
the South Pillar region, generally point toward $\eta$ Carinae and
Tr16 (the lower left of the two stars in Fig.\ 5$b$), as do most
features along the V-shaped dust lane that bisects the nebula.  This
includes a few objects that point toward Tr16, even though they are
projected along the line of sight to Tr14 (see Smith et al.\ 2003a).
Objects within the N bubble, on the other hand, point predominantly
toward Tr14.  Thus, it is clear that Tr16 has more influence on the
southern parts of the nebula, while the younger Tr14 cluster currently
dominates the excavation to the north.\footnote{There are, of course,
some localized exceptions to these general trends due to projection
effects along the line of sight, especially in the brightest inner
regions of the nebula.  A more detailed examination of such
directional indicators and locations of nearby OB stars may illuminate
the role of individual stars in shaping their surroundings on various
size scales, compared to integrated effects from the larger clusters.}

The feature labeled NB in Figure 5$b$ is strange.  It is a nearly
circular shell-like region, but does not appear to contain any bright
OB stars at its center that drive its expansion.  Yet, it has a very
high [S~{\sc ii}]/H$\alpha$ ratio (not shown here, except in the color
image in Fig.\ 1a) that may indicate shock excitation (e.g., Hartigan
et al.\ 1999).  We speculate that NB may mark a position where a
blister is forming; as the main front advances toward northern
Galactic latitudes, it may penetrate through a localized cavity,
injecting hot plasma from the H~{\sc ii} region's interior through the
hole.  Consequently, a blister at the edge of the H~{\sc ii} region
will inflate and the reservoir will eventually pop.  The NB feature
may be a young version of the larger bubble to the south-west (SW in
Fig.\ 5$b$), which appears to have already broken out to larger
distances.  The detailed kinematics of these features have not yet
been investigated.

\subsection{Large-Scale Triggered Star Formation and Carina's Future Prospects}

The South Pillars and other adjacent regions of dense nebular material
around the periphery of the Carina Nebula (Smith et al.\ 2000) show
clear signs of feedback from massive stars at the center of the
nebula, with numerous dust pillars that point inward (Fig.\ 5$b$; see
also Smith et al.\ 2000; Walborn et al.\ 2002; Rathborne et al.\
2002).  The morphology in images (Fig.\ 1) gives the impression that
the remnants of a giant molecular cloud are being overrun and shredded
by the advancing stellar winds and UV radiation from the stars in Tr16
and Tr14 (we do not know what, if anything, triggered the formation of
the first generation clusters Tr14 and Tr16).  Whether or not the
formation of the second generation has been triggered by first
generation feedback is not immediately obvious, because one cannot
determine from morphology alone whether the advancing front triggered
the formation of stars, or if it is simply uncovering dense clumps
that would have formed stars anyway.  However, we explain here how the
scale and character of star formation in Carina may help resolve this
ambiguity.

There are several indications that many of the new second-generation
stars forming in these South Pillars have ages of $\sim$10$^5$ yr or
less.  The clearest evidence so far comes from young massive outflows
and young clusters.  The infamous HH666 jet (Smith et al.\ 2004a) is a
parsec-scale outflow from one of these pillars, showing signs of
powerful bursts of episodic mass loss, hinting at a very early
evolutionary phase undergoing FU~Orionis outbursts (Calvet et al.\
2000).  The embedded class I phase implied by the SED of its driving
source HH666IRS is only expected to last $\sim$10$^5$ yr.  Similarly,
most of the stars in the spectacular ``Treasure Chest'' cluster (Smith
et al.\ 2005b) found in another pillar appear to lie on isochrones
with ages $\sim$10$^5$~yr, and the cluster itself has one of the
highest disk fractions known among any embedded cluster.  There are
several other IR sources and young clusters embedded in pillars in
Carina (Rathborne et al.\ 2002, 2004) that have not been studied as
thoroughly, but are likely to show similar signs of youth.

This recent star formation in the South Pillars is seen across a
region more than half a degree (more than 20 pc) in spatial extent.
The sound crossing time for this region is 18 to 20 Myr or more at a
speed of $c_s\la$1 km s$^{-1}$ in molecular clouds.  In fact,
there are (less vigorous) signs of the youngest phases of star
formation around the entire periphery of the Carina Nebula cavity
across a much larger region of several degrees on the sky (Smith et
al.\ 2000).

The earliest phases of star formation are seen to exist simultaneously
across a region where the sound crossing time is more than an order of
magnitude longer than the duration of these earliest phases.
Therefore, an external agent is required to synchronize these events.
The signal for cloud cores to collapse simultaneoulsy must propagate
at a speed much faster than the sound speed in a molecular cloud.
Since these sources are all seen near the ionized cavity walls and
PDRs of the Carina Nebula, the advancing ionization-shock front is the
likely culprit.  Alternatively, if star formation were constant and
ongoing across the whole Carina Molecular Cloud complex, we might also
expect the youngest population to appear as if it were synchronized
(there would be a range of ages present, which would always include
some fraction of stars in the youngest observable phases).  However,
in that scenario, the stars in the youngest phases should be
distributed randomly about the cloud --- they should not be found
preferentially near the heads of dust pillars that are facing inward
to the massive stars in the core of the nebula, as observed.  Thus, it
is the unique combination of suggestive morphology and the large
spatial scale across which this occurs in Carina that provides a
strong case that these second-generation stars have been directly
triggered.


On a global scale, the relative importance of this triggered second
generation compared to the first generation stars in Tr14 and Tr16 is
not yet clear.  The most massive star currently known in this
second-generation South Pillar region is CPD--59$\arcdeg$2661, which
is an O9 V star at the center of the Treasure Chest cluster (Smith et
al.\ 2005b; Walsh 1984).  The pillar containing the Treasure Chest
cluster is the brightest of the South Pillars at thermal-IR to mm
wavelengths (Smith et al.\ 2000; Rathborne et al.\ 2002; K.\ Brooks et
al., in prep.).  Yonekura et al.\ (2006) have identified a few
additional molecular cores that may be good candidates for sites of
massive star formation.  While obscured regions at the periphery of
Carina may hide additional late O and early B stars (e.g., Sanchawala
et al.\ 2007), it is unlikely that the South Pillars have recently
given birth to large clusters of very massive early O-type stars
comparable to Tr14 and Tr16.  Thus, as star formation cascades to
preferentially lower masses, it would appear to be ``petering out'' in
this second generation that is triggered by stellar winds and
radiation from the first one.  If true, then the current ongoing star
formation is adding mostly low- and intermediate-mass stars to the
fledgeling OB association, but not very massive stars.  The mass
functions in the two regions may be different, but would combine to
form the average mass function of the entire OB association.  Whether
or not this has a significant influence upon the global initial mass
function (IMF) from the region depends on the total mass of stars
added in the second generation, which has not yet been assessed.
Detailed study of this newly-recognized second generation in the South
Pillars is just beginning, but Carina provides a unique laboratory to
investigate this phenomenon.

How long this apparent gentle cascade toward lower-mass star formation
continues into the future is another matter, however.  The massive
object $\eta$ Carinae gives us a constant reminder that the most
massive stars in the region are just now reaching the ends of their
lives.  Its impending demise will be followed by a dozen or so
additional supernovae in the next million years, when the WNL, O3, and
O4 stars in Carina explode (Paper I).  This will suddenly inject
10$^{52}$ to 10$^{53}$ ergs of mechanical energy into the region,
which is likely to sweep up all remaining nebular mass into dense
clouds and trigger further star formation on a massive scale.  In \S 3
we estimated that the available reservoir of nebular mass is at least
1--2$\times$10$^6$ M$_{\odot}$, which is enough to rejuvinate massive
star formation in the region.  At the present time, the majority of
the nebular mass is untapped for star formation because it is in the
atomic gas phase in PDRs that are warmed by UV radiation from the
central clusters.  Less than 30-40\% is in molecular clouds that can
form stars.  In this way, UV emission currently regulates star
formation.

However, after the most luminous stars explode, the UV radiation field
will drop precipitously, because it is the few most massive members
that dominate the FUV luminosity (Paper I).  Their supernova shocks
will sweep this mass into a dense shell that will no longer be warmed
by a strong UV field, and should therefore be more efficient at
forming giant molecular clouds and massive stars.  Essentially, the
disapperance of the UV source plus the shock waves that sweep up the
gas {\it conspire} to enhance the formation of giant molecular clouds.
When this conspiracy unfolds, the new generation of massive star
formation is likely to produce a ring of OB clusters around the fossil
OB association left from Tr14 and Tr16 --- much like the Gould's Belt
and Lindblad ring around the fossil Cas-Tau association in the Solar
neighborhood (Blaau 1991).  Thus, the Carina Nebula may provide us
with a unique snapshot of an earlier phase in the history of our part
of the Galaxy.  Specifically, the current phase of second-generation
star formation in Carina may be directly relevant for understandiing
the origin of young stars near the Sun which reside in the interior of
Gould's Belt.

There is more to this tale as well.  When these multiple supernovae
from $\eta$ Car and its siblings occur, the young stars in the second
generation at the periphery of the Carina Nebula will have their disks
and envelopes pelted by supernova ejecta.  This will happen soon.  The
supernova ejecta will include short-lived radioactive nuclides such as
$^{60}$Fe and $^{26}$Al, which are found in abundance in meteorites in
our Solar System, indicating that the Sun formed in close proximity to
a supernova (Desch \& Ouellete 2005; Hester et al.\ 2004; Tachibana \&
Huss 2003).  An additional clue that the Sun formed close to massive
stars is the clear outer edge of the Kuiper belt (Allen et al.\ 2001;
Jewitt et al.\ 1998), since truncated outer edges are a common
property of externally-evaporated disks in H~{\sc ii} regions, such as
the famous objects seen in silhouette in the Orion Nebula (Bally et
al.\ 2000; McCaughrean et al.\ 1998; McCaughrean \& O'Dell 1996; Smith
et al.\ 2005a).  However, regions like Orion with protoplanetary disks
existing close to just a single young O-type star also have
significant obstacles to acquiring these short-lived radioactive
nuclides.  Namely, while these first-generation stars may be very
close to the supernova, they need to wait 3 Myr or more (about 10 Myr
in the case of Orion) before the massive star explodes, during which
time their gaseous disks will be mostly evaporated.  The stars with
protoplanetary disks that are close enough to $\theta^1$C Ori also
represent a tiny ($\la$1\%) fraction of the surrounding cluster.

These difficulties are alleviated in two important ways in the South
Pillars of Carina and similar regions of ongoing star formation at the
borders of giant H~{\sc ii} regions.  First, this second generation of
stars that has been triggered by the first is located within a few
parsecs of hot massive stars that are already poised to explode, so
there will be {\it no significant delay} between the formation of
these second-generation stars and the arrival of ejecta from the
supernovae.  Because the second generation stars are still young and
have large protoplanetary disks, they are able to cast a wider net to
trap the short-lived radioactive nuclides, as in the ``aerogel'' model
of Desch \& Ouellete (2005).  Second, instead of just one supernova
like in Orion, Carina will have {\it dozens of supernovae} in just a
few million years (there were initially $\sim$70 or more O type stars
in Carina; Paper I).  These two properties make a place like the South
Pillars a much more likely analog of the cradle of the Solar System
than Sun-like stars forming in first generation clusters (like the
protoplanetary disks in the Trapezium of Orion, for example).  The
probability of this happening depends in part on the $^{60}$Fe yield
of SNe from very massive stars like $\eta$~Carinae, which is poorly
constrained (e.g., Smith et al.\ 2007).  It also depends on the
veracity of star formation in regions like the South Pillars, and is
of considerable interest for understanding our own origins.  For this
reason, detailed measurements of the disk fraction, mass function,
total mass, and other properties of stars forming in the South Pillars
is an important goal for future observational studies.


\section{SUMMARY AND CONCLUSIONS}

We have investigated the energy budget and the global properties of
the Carina Nebula on the largest scales of $\sim$3\arcdeg\ (across
100--120 pc).  Furthermore, we have compared these large-scale
properties integrated over the whole nebula with the census of energy
input from OB stars in Paper I (Smith 2006a).  (Note that in all
estimates here, the radiative energy input and the escaping IR
luminosity of the evolved star $\eta$ Carinae have been excluded,
because $\eta$ Car's own circumstellar dust shell processes most of
its bolometric luminosity.)  The main conclusions are summarized here.

1.  The integrated IR luminosity of the Carina Nebula measured in IRAS
    and MSX data is 1.18$\times$10$^7$ L$_{\odot}$.  Dust
    processes about 50--60\% of the known radiative luminosity of the
    OB stars in Carina measured in Paper~I.

2.  From the integrated 3.4 cm radio continuum flux, we infer a total
    Lyman continuum photon luminosity of Q$_H$=6.9$\times$10$^{50}$
    s$^{-1}$.  This is $\sim$75\% of the ionizing photon luminosity we
    derived from a census of Carina's known OB stars in Paper I.  Some
    of the Lyman continuum must therefore escape along with the
    stellar bolometric luminosity or must get absorbed by dust within
    the H~{\sc ii} region.

3.  Using the H$\alpha$ luminosity as a diagnostic of the ionizing
    flux, we measured a smaller value of Q$_H$=3$\times$10$^{50}$
    s$^{-1}$.  This is less than half of that inferred from the radio
    continuum and 1/3 of the total from stars.  The H$\alpha$
    luminosity is significantly absorbed by strong non-uniform
    extinction that is ignored in a standard average reddening
    correction.

4.  Warm dust at $\sim$80~K, which dominates the integrated SED at
    20--30 $\mu$m, resides in the interior of the H~{\sc ii} region
    cavity.  These are probably large grains mixed with ionized
    plasma that are heated {\it in situ} by trapped Ly$\alpha$ from
    the ionized gas.  This warm dust contributes about 1/3 of the
    total IR luminosity, but contains a negligible fraction of the
    total dust mass.  Based on the similarity of the radio continuum
    and MSX Band-E images, we find that diffuse 20-25 $\mu$m emission
    is an excellent tracer of warm dust mixed with ionized gas inside
    H~{\sc ii} regions.

5.  Cooler dust at 30--40~K dominates the far-IR emission, the total
    IR luminosity, and the total dust mass of the nebula.  From the
    35~K component fit to the SED, we measure a total dust mass of
    $\sim$10$^4$ M$_{\odot}$, indicating a likely gas mass associated
    with this dust of 10$^6$ M$_{\odot}$ or more.

6.  With this total mass and the ubiquitous $\pm$20 km s$^{-1}$
    expansion seen across the nebula, we infer a total kinetic energy
    in moving gas of at least 8$\times$10$^{51}$ ergs.  This is only
    30\% of the available mechanical energy from stellar winds during
    the 3 Myr lifetime of the nebula.  Therefore, on energy grounds,
    there is no need to invoke a previous SN to explain the kinematics
    of the region.

7.  We find a near-perfect correlation between the 60 $\mu$m optical
    depth and PAH emission seen in the MSX band-A image.  The only
    violations of this correlation are point sources with warm
    circumstellar dust.  The PAH emission and 60 $\mu$m optical depth
    have a different spatial distribution than the largest molecular
    clouds traced by CO emission.  The total gas mass traced by
    $\sim$35~K dust in PDRs outweighs the total gas mass in molecular
    clouds.  Since these two components have different spatial
    distributions, we conclude that most of the mass in the nebula
    resides in atomic gas in PDRs instead of dense molecular clouds,
    and is therefore not currently participating in star formation.
    However, the atomic and molecular gas added together provide a
    substantial reservoir of 1--2$\times$10$^6$ M$_{\odot}$ for future
    star formation.

8.  The large sound crossing time of several Myr for the South Pillars
    is much longer than the $\sim$10$^5$ yr duration of the earliest
    phases of star formation, seen to be synchronized throughout the
    region.  This provides a strong case that much of the ongoing star
    formation in Carina is indeed triggered, instead of just
    occurring spontaneously and then being uncovered by the advancing
    ionization front.

9.  The triggered second-generation of star formation appears to lack
    the same scale of high-mass star formation that gave rise to the
    massive first-generation clusters like Tr14 and Tr16.  This
    implies that star formation triggered by stellar winds and UV
    radiation has a cascading effect, biased toward lower masses.
    Determinig if this is indeed true and how it affects the IMF will
    be an important aspect of future research in this region.  We
    suspect that the large number of sequential supernovae to occur in
    the near future may conspire with the lack of FUV radiation from
    those same stars to rejuvinate {\it massive} star formation, given
    the large reservoir of atomic and molecular mass still available.

10.  Regions like the South Pillars, where a second generation of
    young stars has been triggered at the periphery of a giant H~{\sc
    ii} region, will soon be bombarded with supernova ejecta.  We
    argue that these are the best candidates for analogs of the type
    of environment where the Sun formed, given current constraints
    from meteoritic evidence of short-lived radioactive nuclides.
    Dozens of sequential supernovae in a short time plus the lack of a
    delay time between their birth and the arrival of the supernova
    ejecta allow for more efficient injection of short-lived
    radioactive nuclides in these second-generation triggered regions,
    as compared to environments like the Trapezium in Orion.

\smallskip\smallskip\smallskip\smallskip
\noindent {\bf ACKNOWLEDGMENTS}
\smallskip
\scriptsize

We thank the ATNF Distinguished Visitor Program, providing us with an
opportunity to collaborate in person in Australia.  We acknowledge
fruitful discussions with John Bally, Hans Zinnecker, Ed Churchwell,
Gus Muench, Norm Murray, and Jonathan Williams.  Partial support was
provided by NASA through grant HF-01166.01A from the Space Telescope
Science Institute, which is operated by the Association of
Universities for Research in Astronomy, Inc., under NASA contract
NAS~5-26555.

\end{document}